\documentclass[twocolumn]{aastex701}
\usepackage{enumitem}
\usepackage{amsmath}
\usepackage{booktabs}
\newcommand{\p}[1]{\textbf{{\color{blue}{#1}}}}

\shorttitle{Solar active region detection and tracking}
\shortauthors{C. X. Shi et al.}

\begin{document}

%\title{HARDAT: HDBSCAN based Solar Active Region Detection and Tracking}
\title{An Improved HDBSCAN-based Detection and Tracking Method for Solar Active Regions in Magnetograms}
\correspondingauthor{Q. Hao}

\author[0009-0005-2415-4084]{C. X. Shi}
\affiliation{School of Astronomy and Space Science, Nanjing
	University, Nanjing 210023, People's Republic of China}
\email{scx18227932597@163.com}

\author[0000-0002-9264-6698]{Q. Hao}
\affiliation{School of Astronomy and Space Science, Nanjing
	University, Nanjing 210023, People's Republic of China}
\affiliation{Key Laboratory of Modern Astronomy and Astrophysics
	(Nanjing University), Ministry of Education, Nanjing 210023, People's Republic of China}
\email[show]{haoqi@nju.edu.cn}

\author[0000-0002-7289-642X]{P. F. Chen}
\affiliation{School of Astronomy and Space Science, Nanjing
	University, Nanjing 210023, People's Republic of China}
\affiliation{Key Laboratory of Modern Astronomy and Astrophysics
	(Nanjing University), Ministry of Education, Nanjing 210023, People's Republic of China}
\email{chenpf@nju.edu.cn}

\author[0000-0002-9293-8439]{Y. Guo}
\affiliation{School of Astronomy and Space Science, Nanjing
	University, Nanjing 210023, People's Republic of China}
\affiliation{Key Laboratory of Modern Astronomy and Astrophysics
	(Nanjing University), Ministry of Education, Nanjing 210023, People's Republic of China}
\email{guoyang@nju.edu.cn}

\begin{abstract}
Solar active regions (ARs) are the primary source of solar eruptions and space weather. Accurate detection and tracking of ARs is crucial for understanding their evolution and predicting solar activities. In the previous work, based on the density-based spatial clustering of applications with noise (DBSCAN) approach, we proposed the DBSCAN-based solar active region detection (DSARD) framework. To overtake its limitations, in this paper we applied the hierarchical density-based spatial clustering of applications with noise (HDBSCAN) approach to the detection of solar active regions, which is called the HDBSCAN-based solar active region detection and tracking (HARDAT) method. This enables the algorithm to handle multi-density magnetic structures dynamically, eliminating the need for fixed thresholds. Consequently, the algorithm can detect diffuse and small ARs more effectively while preserving morphological integrity. We have also developed a solar differential rotation based tracking algorithm that integrates physical motion models and Hamming distance similarity metrics to achieve robust multi-object tracking. Additionally, we propose a novel polarity inversion line extraction method that uses support vector classification, which offers superior generalization for complex AR boundaries. Processing line-of-sight magnetograms from SOHO/MDI (1996--2011) and SDO/HMI (2010--2024) and evaluating them against the National Oceanic and Atmospheric Administration (NOAA) and DSARD catalogues demonstrates that HARDAT is superior in terms of sensitivity, accuracy, and stability of detection and tracking. This is particularly evident when resolving clustered ARs and maintaining identity continuity. HARDAT therefore offers a comprehensive solution for the long-term analysis of AR evolution and space weather prediction.
\end{abstract}

\keywords{\uat{The Sun}{1693} --- \uat{Solar active regions}{1974} --- \uat{Solar magnetic fields}{1503} --- \uat{Solar cycle}{1487} --- \uat{Astronomical techniques}{1684} }

\section{Introduction} \label{sec:intro}
Solar active regions (ARs) are localized areas of strong magnetic fields on the Sun. They are often associated with sunspots, flares and coronal mass ejections (CMEs; \citealt{van2015evolution}). Research on ARs involves analysing magnetic field structures, their evolution and the mechanisms by which energy accumulates within these regions, to assess the potential risk of eruptive events, such as solar flares and CMEs \citep{chen11}. Studying them is essential for understanding solar dynamics and forecasting space weather \citep{Toriumi2019}. High-resolution and high-cadence magnetographs, such as the Michelson Doppler Imager (MDI; \citealt{Scherrer1995}) on board the Solar and Heliospheric Observatory (SOHO; \citealt{Domingo1995}) and the Helioseismic and Magnetic Imager (HMI; \citealt{Schou2012}) on board the Solar Dynamics Observatory (SDO; \citealt{Pesnell2012}), have rendered traditional processing methods inadequate for handling their massive datasets. These methods rely on manual inspection or threshold-based image processing. 

 Consequently, researchers have developed a series of automated detection methods for ARs in magnetograms \citep{Zhang2010,Turmon2010,Bobra2014,Quan2021,Jiang2022,Wang2023,chen2025statistical}. These methods provide crucial data support for the studies of solar ARs, solar activities, long-term solar dynamo modeling. In particular, we recently applied the unsupervised desity-based spatial clustering of applications with noise (DBSCAN; \citealt{Schubert2017}) algorithm to develop the DBSCAN-based solar active region detection (DSARD; \citealt{chen2025statistical}) method. DSARD eliminates the need for the vast high-quality annotated data that are inherent in supervised learning, thereby reducing subjective bias. Its density-based nature enables the adaptive detection of ARs across varying intensities, including small-scale transient magnetic features that are often overlooked by conventional thresholding methods. This significantly enhances the accuracy and robustness of detection. Employing staged clustering in a two-stage DBSCAN density clustering process effectively avoids the tendency of the traditional methods (such as region growing and morphological algorithms) to excessively merge adjacent ARs or perform inappropriate segmentation within high-resolution magnetograms. This significantly improves the capability to analyze fine-scale magnetic structures.  Compared with the National Oceanic and Atmospheric Administration (NOAA) and Spaceweather HMI Active Region Patch (SHARP; \citealt{Bobra2014}), the DSARD  method achieved accuracy rates of $89.7\%$ and $78.2\%$, respectively, whilst identifying small-scale active regions that are overlooked by conventional methods. 

We conducted detailed statistical analyses of characteristics such as the number, area, magnetic flux, tilt angle, and drift velocity of active regions during the solar cycle 24 and the rising phases of the solar cycle 25 using the active region database generated by the DSARD method. The majority of the statistical results were consistent with those of previous studies, thereby validating the reliability of the DSARD method. Furthermore, our research yielded several novel key conclusions such as the finding that large ARs adhere to Joy's Law, whereas small ARs exhibit random distribution patterns. The proportion of ARs that violate Hale's Law is markedly higher than in previous studies. 

However, we also found that the DSARD method had some limitations. This is because the DBSCAN algorithm uses fixed density threshold and distance parameter, which restricts its ability to detect diffuse regions and multi-density magnetic field structures. Furthermore, DSARD does not track the detected ARs. To avoid duplicate detections and minimise the projection effects, a margin of $\pm 6^{\circ}$ from the center meridian of the solar disk is selected when evaluating the detection results. Consequently, the true positive rate of the results is therefore limited.  

Hierarchical density-based spatial clustering of applications with noise (HDBSCAN; \citealt{mcinnes2017hdbscan,campello2013density}) is a powerful extension of DBSCAN that can handle clusters with varying densities by leveraging the concepts of mutual reachability distance and cluster stability. In this paper, we propose an HDBSCAN-based solar active region detection and tracking (HARDAT) method. The adoption of HDBSCAN makes it particularly suitable for magnetograms, where ARs exhibit a wide range of sizes and magnetic density distributions. Our method improves upon DSARD by incorporating dynamic density clustering and a tracking module informed by physics that accounts for solar differential rotation \citep{schroter1985solar, Rao2024}. We also propose a new polarity inversion line (PIL) extraction method based on support vector classification (SVC; \citealt{708428}). This method has a stronger generalization ability than the traditional mask-based method and can simultaneously extract magnetic neutral lines in dense and diffuse magnetic fields. The paper is organized as follows: Section~\ref{data acquisition} describes the data used in this study. Section~\ref{detection} details the improved detection methodology, including HDBSCAN clustering and integration strategies. Section~\ref{track} presents the tracking algorithm based on solar differential rotation. Section~\ref{analysis} introduces the SVC-based PIL extraction method. Section~\ref{discussion and conclusion} provides a discussion and conclusion, highlighting the advantages of HARDAT over existing methods.

\section{Data Acquisition} \label{data acquisition}
This study uses the solar full disk line-of-sight (LOS) magnetograms, which were provided by SOHO/MDI from the year 1996 to 2011 and SDO/HMI from 2010 to 2024. The SOHO/MDI magnetogram dataset provides solar photospheric magnetic distribution with a spatial resolution of 4\arcsec\ (an image size of 1024×1024 pixels). SDO/HMI, the successor to SOHO/MDI, has greater temporal and spatial accuracy. It provides solar magnetograms with a spatial resolution of 1\arcsec\ (an image size of 4096×4096 pixels). And we use $mdi.fd\_M\_96m\_lev182$ and $hmi.M\_720s$ magnetorgrams data series for this study. As the cadences of these two datasets are differ, the sampling cadence of the data is set to 192 minutes for SOHO/MDI magnetograms and 3 hours for SDO/HMI magnetograms,  ensuring the consistency in subsequent tracking accuracy. This yields solar magnetogram data spanning a total of approximately 28 years, from 1996 to 2024, including the whole solar cycles 23 and 24, and the rising phase of solar cycle 25. All data for this study are retrieved from the Joint Science Operations Center of Stanford\footnote{\url{http://jsoc.stanford.edu}}.

\section{Detection} \label{detection}

The original DSARD method consists of the following four steps: (1) threshold-based image segmentation, (2) first global DBSCAN, (3) second reformative DBSCAN, and (4) integration. The corresponding parameters are described in Table~\ref{tab: DSARD}, except for $\beta$ and $\gamma$. First, it extracts seed pixels with magnetic field strengths exceeding the $threshold$. It then applies two iterations of DBSCAN clustering in the ($x$, $y$) space to group adjacent pixels using the parameters $\epsilon$, $minSamples\_1$, and $minSamples\_2$. Subsequently, the resulting cluster types are categorized according to the ratio of positive to negative polarity pixels. The distances between different clusters are then calculated based on their types and compared with $minDistance$ to determine whether these clusters should be merged. Finally, the area of each cluster is calculated and filtered using $minSize$ to obtain the final result. Since the clustering results of DBSCAN are mainly affected by two parameters, $\epsilon$ and $minDistance$, which are actually equivalent to a fixed density threshold, the generalization ability of DBSCAN is not strong when used alone. The advantage of DSARD lies in its application of two iterations of DBSCAN to address the problem of uneven density distribution within the strong-field regions of active regions. This approach mitigates the limitations of DBSCAN to a certain extent. However, the model results remain highly dependent on parameter selection, and the use of two iterations of DBSCAN procedures effectively provides only two density thresholds, thereby failing to achieve clustering with adaptive density thresholds. The same issue also arises in the integration step, where the merging criterion is a fixed parameter, $minDistance$, which cannot accommodate most cases. Therefore, this section focuses on optimizing the integration step in the DSARD model and proposes a new model based on HDBSCAN.

\subsection{Modification of the DSARD Method}  \label{DSARD}

The original DSARD integration algorithm merges clusters based on distance and cluster type. This process is mainly determined by the parameters $minDistance$ and $ratio$. However, the former parameter is derived from experience. It has a fixed value, and is independent of sample data and clustering results. Therefore, applying the same set of integration algorithms to the clusters obtained after two iterations of DBSCAN clustering may not be optimal, potentially resulting in suboptimal morphological outcomes in the detected ARs. Figure~\ref{fig: DSARD_disadvantage_1}(a) shows the magnetogram in 2000 May 20, and panel (b) and (c) are the results of first and second DBSCAN clustering, respectively. And in panel (d), AR2 and AR12, which should be considered as one AR, are splited due to a small $minDistance$. While the parameter settings in the example differ from those of the original DSARD method, they indicate the potential drawback with the original DSARD method. To address this issue,  we introduce a term that takes cluster size into account in the distance calculation, and introduce two parameters $\beta$ and $\gamma$ to adjust the influence of this term. This allows us to calculate the distance between clusters more dynamically, thereby improving the robustness of the model. The specific implementation process is as follows:

\begin{figure*}
    \centering
    \includegraphics[width=\linewidth]{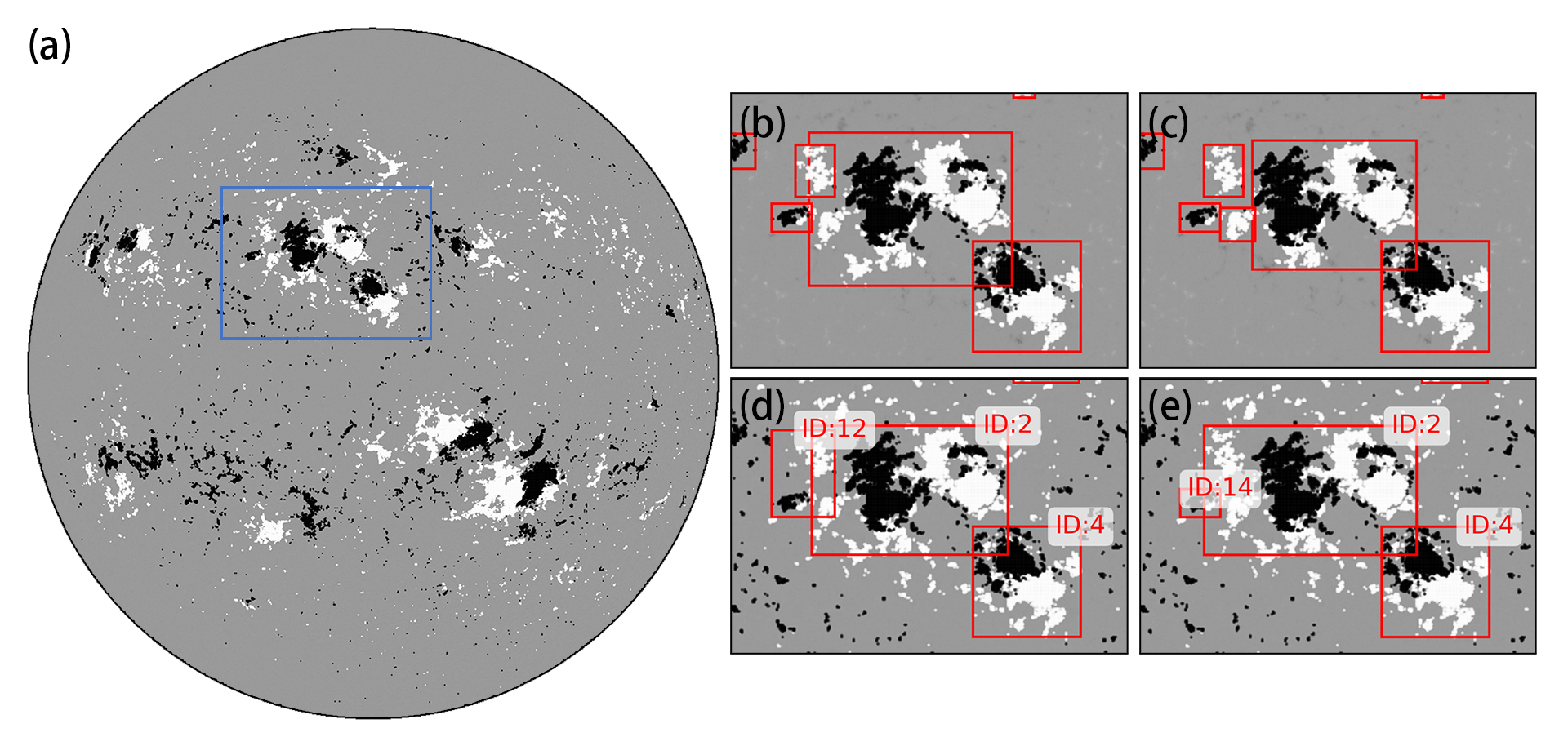}
    \caption{The DSARD result for the SOHO/MDI magnetogram on 2000 May 20, with $minDistance$ set to 20 Mm and $minSize$ set to 70 Mm$^2$. Panel (a) The result after threshold segmentation. The blue box area is the target region. The right panels are the result of the target region after each part of DSARD processing. Panel (b) Clustering results of the first global DBSCAN. Panel (c) Clustering results of the second reformative DBSCAN. Panel (d) Consequence of integration. Panel (e) Better integration with $\beta$ set to 0.5 and $\gamma$ set to 0.5.}
    \label{fig: DSARD_disadvantage_1}
\end{figure*}

\begin{enumerate}[leftmargin=*]
    \item In the original DSARD method, the nearest distance between two clusters after the second DBSCAN is calculated according to the $cKDTree$ algorithm \citep{virtanen2020scipy}. Here, we modify the method by sorting the distances and selecting $95\%$ of them as the nearest distance, in order to reduce the influence of outliers on the calculation distances.

    \item The corrected distance is then obtained by multiplying the nearest distance by the coefficient $k$, consisting of three terms. The first term is determined by the types of the two clusters and is the parameter $\alpha$ in the original DSARD method. $\alpha$ is a coefficient determined by the cluster categories to define a measurement of the distance between clusters (see Table 1 in \citet{chen2025statistical}). The second and third terms are the newly added correction terms and are determined by the sizes of two clusters and the external parameters $\beta$ and $\gamma$. The formula is as follows:
        \begin{equation}
            A_{\text{sum}} = A_1 + A_2, r = \frac{\max(A_1, A_2)}{\min(A_1, A_2)},
            \label{eq: eq1}
        \end{equation}
        \begin{equation}
            k_1 = \alpha,
            \label{eq: eq2}
        \end{equation}
        \begin{equation}
            k_2 = \beta \frac{A_{\text{sum}}}{A_{\text{sum}}+A_0} (1 - \frac{r - 1}{r + 1}),
            \label{eq: eq3}
        \end{equation}
        \begin{equation}
            k_3 = \gamma \frac{r - 1}{r + 1},
            \label{eq: eq4}
        \end{equation}
        \begin{equation}
            k = k_1 max(0.8, min(1 + k_2 - k_3, 1.2)),
            \label{eq: eq5}
        \end{equation}
    where $A_1$ and $A_2$ represent the areas of the two clusters after projection correction. $A_0$ represents 0.5 times the area of the largest cluster after the two DBSCAN clustering iterations of the current magnetogram. $A_0$ is used as the standard by which the cluster size is measured. The $k_2$ term in Equation~(\ref{eq: eq3}) indicates that the distance between two large clusters increases more and the distance between two small clusters or between one large and one small cluster increases less. This reduces the probability of two adjacent large active regions merging into one. The $k_3$ term in Equation~(\ref{eq: eq4}) aims to minimize the distance between large and small clusters. This is motivated by the diffuse nature of the AR's edge and the density-based characteristics of the DBSCAN algorithm, meaning that the clustering results occasionally exhibit a large cluster surrounded by numerous smaller ones. These two terms $k_2$ and $k_3$ can facilitate the merging of nearby smaller clusters and enhance the structural integrity of the active region. With the modified method, the magnetogram on 2000 May 20 is processed again, and the identification of ARs is displayed in  Figure~\ref{fig: DSARD_disadvantage_1}(e), which shows the result of better integration, i.e., AR2 has a more complete form than that in Figure~\ref{fig: DSARD_disadvantage_1}(d). A small AR14 is not included in AR2 because it represents a bipolar region, and the parameter $\alpha$ causes it not to merge with AR2. 
    
    Despite the improvement, the inclusion of parameters $\beta$ and $\gamma$ is more appropriate for adjustment in individual case studies, and is not ideal for automatically detecting large samples. This is because, in certain scenarios such as when multiple small clusters exist between two active regions, these clusters may be combined as a consequence of these parameters, which would lead to an unintended outcome. A specific analysis of these two parameters can be found in Appendix~\ref{param_analysis}. Therefore, setting these parameters to zero is preferable for automated detection in statistical studies.
    
\end{enumerate}

\subsection{HDBSCAN-based Solar Active Region Detection (HARD)} \label{HARD}

\begin{figure*}
    \centering
    \includegraphics[width=\linewidth]{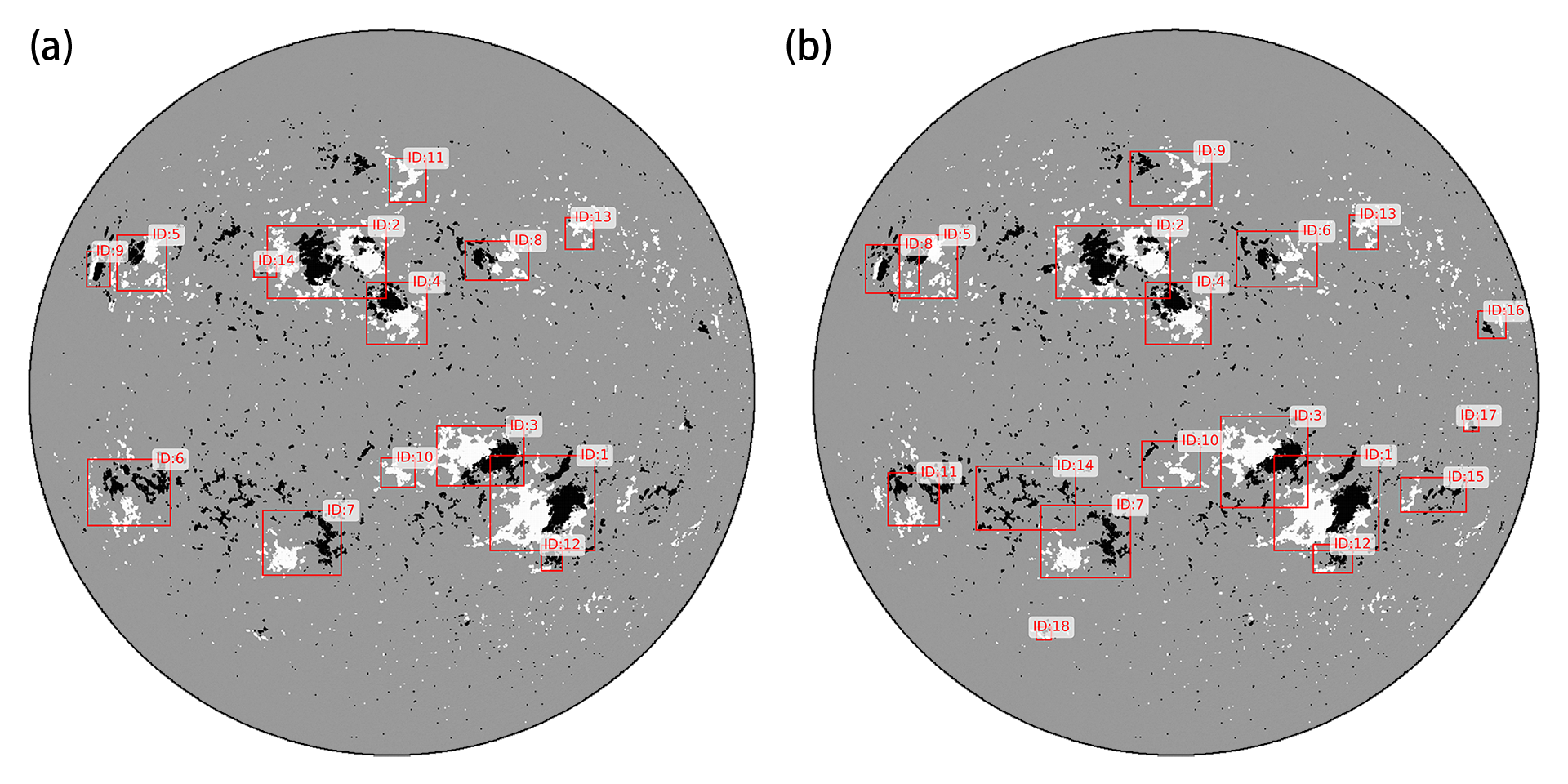}
    \caption{Panel (a) is the result under the parameters in Figure~\ref{fig: DSARD_disadvantage_1}. Panel (b) is the result of setting the $minSamples\_1$ to 60.}
    \label{fig: DSARD_disadvantage_2}
\end{figure*}

The DBSCAN clustering algorithm relies on a fixed density threshold, which may limit its performance when dealing with datasets that exhibit complex or varying density distributions. While the DSARD method employs a two iterations of DBSCAN approach to address this challenge and detect active regions with different density characteristics, the generalization capability of this strategy could be further improved in practical applications. Comparing Figures~\ref{fig: DSARD_disadvantage_2}(a) and (b), the density threshold of the first global DBSCAN clustering in panel (a) is too high, resulting in the failure to detect the small active regions including AR16, AR17, AR18 and the dispersion such as AR15 detected in panel (b). At the same time, the morphology of the active regions in Figure~\ref{fig: DSARD_disadvantage_2}(a) with AR8, AR11 also has the disadvantage of being incomplete. While it is possible to adjust the parameters of the first global DBSCAN to alter the density threshold and detect more small or diffuse structures, as shown in Figure~\ref{fig: DSARD_disadvantage_2}(b), this will cause the originally detected regions to expand until they exceed the artificially set area threshold $maxSize$, resulting in the second DBSCAN clustering with high density threshold and the destruction of the structure, as in AR11 in Figure~\ref{fig: DSARD_disadvantage_2}(b).

The purpose of the two iterations of DBSCAN is to address the multi-density distribution. There are corresponding algorithms in the field of clustering algorithms, such as ordering points to identify the clustering structure (OPTICS; \citealt{ankerst1999optics}) and HDBSCAN, which are derivative algorithms of the DBSCAN algorithm for clustering dynamic density data. Therefore, the effect of the two-iteration DBSCAN method in DSARD can be implemented by using these advanced clustering algorithms. After some experiments about these advanced algorithms, HDBSCAN is found to be a promising candidate.

HDBSCAN introduces two new concepts based on DBSCAN, which are \textbf{core distance} and \textbf{mutual reachability distance}: 
\begin{itemize}
  \item \textbf{Core distance:} The shortest distance that makes point $P$ the core point, i.e., the distance from the nearest $min\_samples$ - 1 point to $P$.
  \item \textbf{Mutual reachability distance:} The maximum among the distance between $P1$ and $P2$, the core distance of $P1$ and the core distance of $P2$.
\end{itemize}
The distance matrix is constructed using mutual reachability distance, which keeps the dense points (with low core distance) at a similar distance from each other and extends the distance between diffuse points (with high core distance). This ultimately results in an obvious difference in the density of the point set. A distance matrix is then obtained by calculating the mutual reachability distance between points. A hierarchical structure is obtained by constructing a minimum spanning tree based on the distance matrix \citep{cheriton1976finding}. Next, the tree is simplified using the parameter $minClusterSize$, and the stability of the clusters is defined to separate or merge them. Finally, dynamic density clustering results are obtained without relying on a density threshold.

\begin{figure*}
    \centering
    \includegraphics[width=\linewidth]{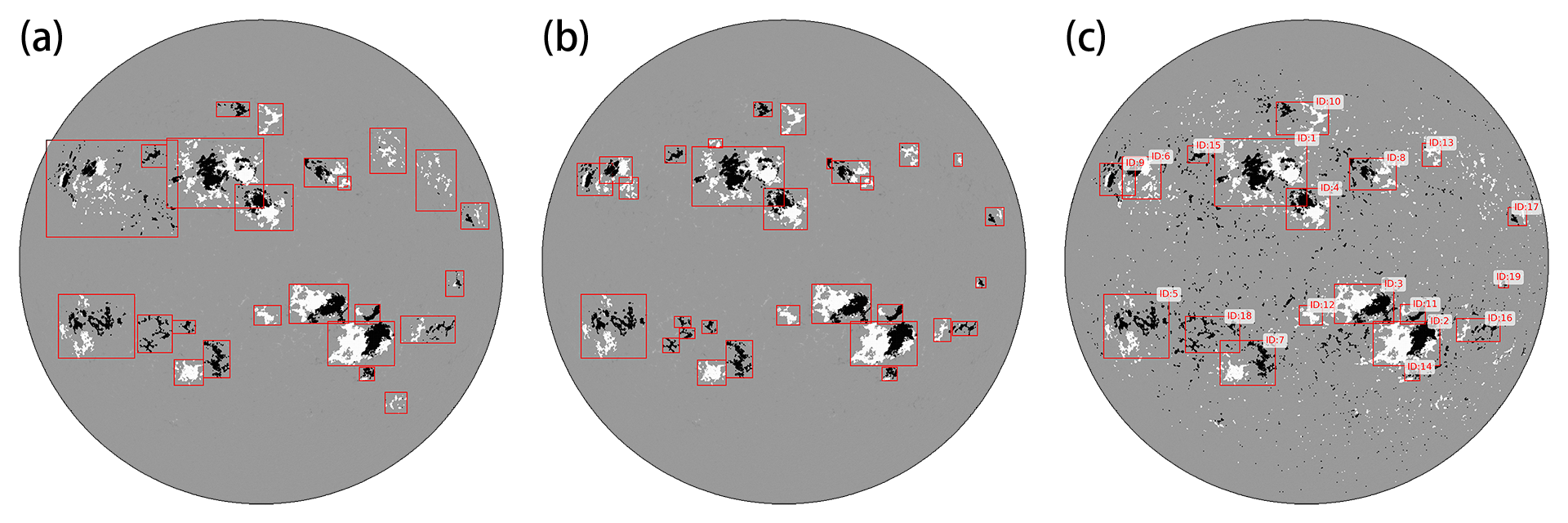}
    \caption{HDBSCAN-based solar active region detection result of the SOHO/MDI magnetogram on 2000 May 20. Panel (a) HDBSCAN clustering results. Panel (b) DBSCAN clustering results. Panel (c) Better integration with $\beta$ set to 0.5 and $\gamma$ set to 0.5. }
    \label{fig: HDBSCAN}
\end{figure*}

Figure~\ref{fig: HDBSCAN} shows the process of the AR detection algorithm after HDBSCAN is introduced. The aim is to use HDBSCAN to achieve the two iterations effect of DBSCAN directly, so HDBSCAN clustering is employed immediatedly after the magnetic field strength threshold segmentation. Figure~\ref{fig: HDBSCAN}(a) shows the result of HDBSCAN clustering. It can be seen that the two iterations effect of DBSCAN is achieved and that small structures and diffuse regions in the western region of the solar disk are detected, as well as the distinction between adjacent large structures. However, HDBSCAN algorithm does not remove the noise in the northeast region well because it is classified as one cluster in the HDBSCAN stability algorithm. Therefore, DBSCAN with a small density threshold is performed again in Figure~\ref{fig: HDBSCAN}(b), to remove the noise in the region. The small density threshold is chosen to prevent the damage to the structure in the detected region. The final results are displayed in Figure~\ref{fig: HDBSCAN}(c). Comparing Figure~\ref{fig: DSARD_disadvantage_2} with Figure~\ref{fig: HDBSCAN}(c), it is seen that introducing HDBSCAN achieves smaller, more diffuse region detection while retaining the morphological integrity of the active region. However, this advantage comes at the cost of increased performance overhead. More details about the HDBSCAN clustering on solar magnetograms can be seen in Appendix~\ref{HDBSCAN flowchart}.

\subsection{Performance of The Detection Methods} \label{detection performance}

Both the DSARD method based on DBSCAN and the HARD method based on HDBSCAN are used for active region detection in SOHO/MDI and SDO/HMI magnetograms. However, due to the different spatial resolutions of these two types of magnetograms, the parameters selected even with the same model are different for the two data sets. The parameters used in this study are shown in Table~\ref{tab: DSARD} and Table~\ref{tab: HDBSCAN}. 

\begin{table}[h!]
  \begin{center}
    \caption{Parameters in the DSARD model.}
    \begin{tabular}{l|c|c} 
      \hline
      \textbf{Parameter} & \textbf{SOHO/MDI} & \textbf{SDO/HMI} \\
      \hline
      $threshold$ & 100 Gauss & 150 Gauss \\
      $\epsilon$ & 10 pixels & 30 pixels\\
      $minSamples\_1$ & 100 pixels & 200 pixels\\
      $minSamples\_2$ & 150 pixels & 500 pixels\\
      $maxSize$ & 3000 Mm$^2$ & 3000 Mm$^2$\\
      $ratio$ & 10 & 10\\
      $minDistance$ & 40 Mm & 40 Mm\\
      $minSize$ & 70 Mm$^2$ & 70 Mm$^2$\\
      $\beta$ & 0--1 (default 0) & 0--1 (default 0)\\
      $\gamma$ & 0--1 (default 0) & 0--1 (default 0)\\
    \hline
    \end{tabular}\label{tab: DSARD}
  \end{center}
\end{table}

\begin{table}[h!]
  \begin{center}
    \caption{Parameters in the HARD model.}
    \begin{tabular}{l|c|c} 
      \hline
      \textbf{Parameter} & \textbf{SOHO/MDI} & \textbf{SDO/HMI} \\
      \hline
      $threshold$ & 100 Gauss & 150 Gauss \\
      $minClusterSize$ & 100 pixels & 500 pixels\\
      $minSamples\_H$ & $None$ & 100 pixels\\
      $\epsilon\_ClusterSelection$ & $None$ & 30 pixels\\
      $\epsilon$ & 10 pixels & 40 pixels \\
      $minSamples$ & 50 pixels & 100 pixels \\
      $ratio$ & 10 & 10\\
      $minDistance$ & 40 Mm & 40 Mm\\
      $minSize$ & 70 Mm$^2$ & 70 Mm$^2$\\
      $\beta$ & 0--1 (default 0) & 0--1 (default 0)\\
      $\gamma$ & 0--1 (default 0) & 0--1 (default 0)\\
    \hline
    \end{tabular}\label{tab: HDBSCAN}
  \end{center}
\end{table}

In Table~\ref{tab: HDBSCAN}, the $minClusterSize$, $minSamples\_H$ and $\epsilon\_ClusterSelection$ are the HDBSCAN parameters, while the $\epsilon$ and $minSamples$ are the DBSCAN parameters. In the analysis of SOHO/MDI magnetograms, the parameters $minSamples\_H$ and $\epsilon\_ClusterSelection$ are set to ``None". This contrasts with the processing of SDO/HMI magnetograms, where these parameters are given specific values. This discrepancy arises from the lower spatial resolution in SOHO/MDI magnetograms, which results in a simpler data distribution. This allows for effective outcomes with a single parameter configuration. Conversely, the higher spatial resolution of SDO/HMI magnetograms results in a more intricate data distribution. This necessitates adjusting two additional parameters, $minSamples\_H$ and $\epsilon\_ClusterSelection$, to achieve optimal results. 

\begin{figure*}
    \centering
    \includegraphics[width=\linewidth]{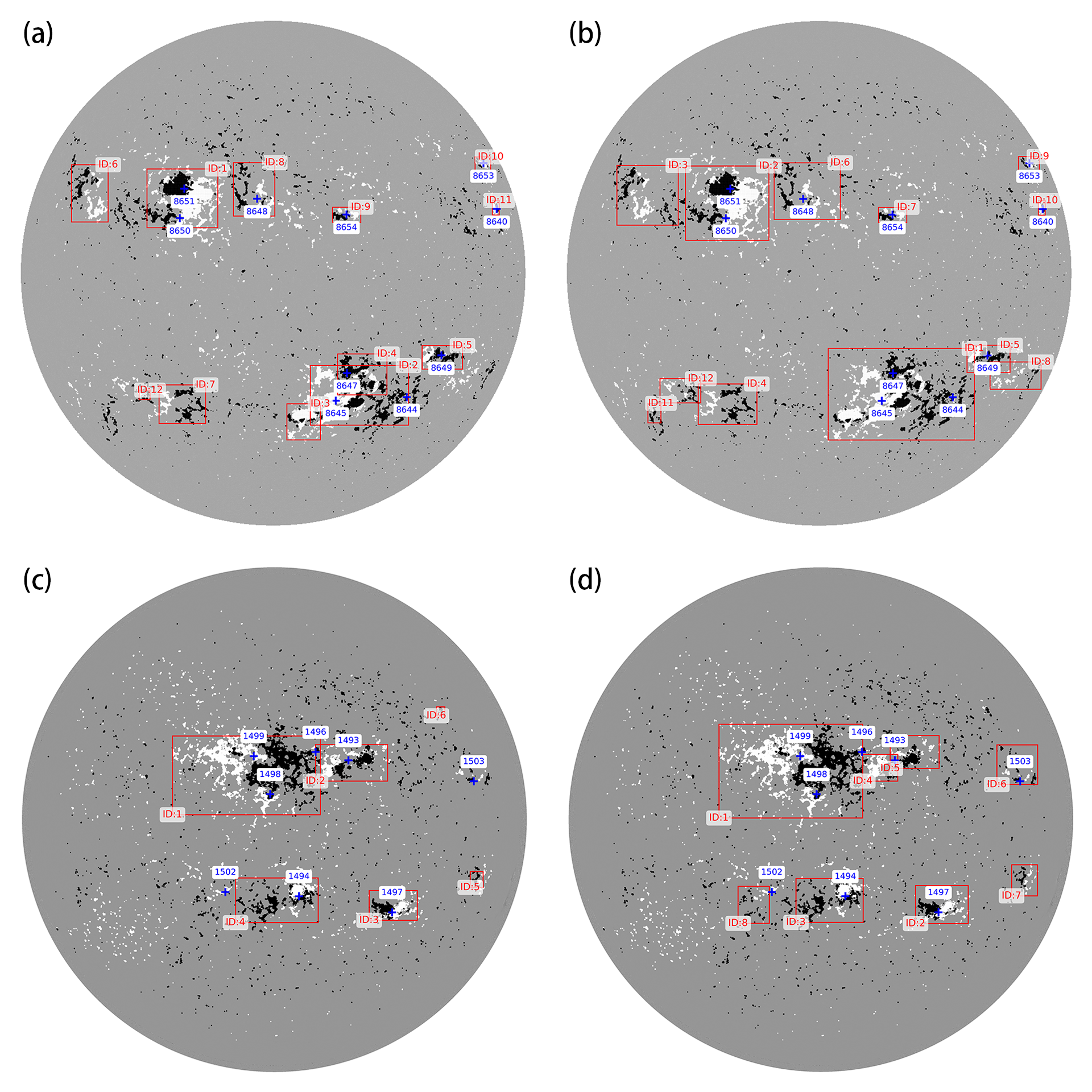}
    \caption{Solar active region detection results of the SOHO/MDI magnetogram on 1999 August 1 (a, b) and 2012 June 7 (c, d). Each red box with a number represents a bounding rectangle of a solar active region identified by DSARD (a, c) or HARD (b, d). The blue plus signs show the labeled ARs of NOAA.}
    \label{fig: DSARD_HDBSCAN_NOAA}
\end{figure*}

Figure~\ref{fig: DSARD_HDBSCAN_NOAA} shows the detection results of DSARD and HARDAT on MDI and HMI magnetograms, respectively. The blue plus signs show the labeled ARs of NOAA. DSARD and HARD demonstrate complementary performance characteristics across various samples. In Figure~\ref{fig: DSARD_HDBSCAN_NOAA}(a), DSARD identifies three active regions, AR2, AR3 and AR4, whereas in Figure~\ref{fig: DSARD_HDBSCAN_NOAA}(b), HARD identifies these ARs as the same AR1. Conversely, the distinct active regions AR4 and AR5 depicted in Figure~\ref{fig: DSARD_HDBSCAN_NOAA}(d) are not separable in Figure~\ref{fig: DSARD_HDBSCAN_NOAA}(c). To objectively assess the advantages and disadvantages of DSARD and HARD, we chose the labeling outcome provided by NOAA as the ground truth. An accurate detection is defined as when the center locations reported by NOAA fall within the minimum bounding rectangle of the detected ARs, which is consistent with previous works \citep{Zhang2010,chen2025statistical}. After tracking the detection results over time in Section~\ref{track}, we quantitatively calculated assessment metrics for comparison.

\section{Tracking} \label{track}

Due to the dynamic nature of solar active regions, which encompasses processes such as merging, splitting, and migration, long-term monitoring and evolutionary analysis of these regions is essential. We made the improvement of the DSARD method for detecting solar active regions and introduced a new automated detection method based on HDBSCAN, called HARD. However, the detected active regions are not tracked or numbered. Therefore, in this section we implement tracking of solar active regions based on the DSARD and HARD detection methods.

The process of tracking solar active regions can be attributed to a multiple object tracking (MOT) problem \citep{luo2021multiple}, and the classic approach is the tracking-by-detection method. The fundamental concept behind this technique is data association, meaning that the targets detected in the current frame are matched with the pre-existing targets to enable continuous tracking. Despite the existence of several well-established MOT algorithms, such as Simple Online and Realtime Tracking (SORT, \citealt{bewley2016simple}) and Channel and Spatial Reliability Tracking (CSRT, \citealt{lunevzivc2018discriminative}), tracking solar active regions poses distinct challenges. These challenges arise from various issues including morphological transformations, projection effects, and variable sampling time interval precision, all of which diminish the performance of conventional tracking techniques when applied to solar active regions. For example, SORT tracking presupposes relatively uniform target motion. However, this assumption can lead to tracking errors when the sample time intervals are variant. To address these challenges, the tracking algorithm should integrate the solar differential rotation to predict the location of active regions. By utilizing the predictable movements of the solar photosphere surface, this technique improves the tracking accuracy over time. This method ensures that active regions are accurately identified and numbered, thereby supporting more robust and reliable monitoring of solar activity.

\subsection{Solar Differential Rotation based Active Region Track} \label{track method}

The solar differential rotation is characterized by varying angular velocity across different latitudes, driven by internal convection and the Coriolis effect, resulting in higher angular velocity at lower latitudes and reduced angular velocity at higher latitudes \citep{Schou1998,Thompson2003,Rao2024}. This variability helps determine the location and general morphological features of an active region accurately after a certain period, ignoring its evolutionary changes. However, it is important to consider the evolution of active region actually. Thus, enhancing the temporal resolution of the sample and ensuring the real-time updates of the target are necessary. This investigation employed a temporal sampling resolution of 192 minutes for SOHO/MDI magnetograms and three hours for SDO/HMI magnetograms, as detailed in Section~\ref{data acquisition}. The comprehensive method used in this study to track the detected active regions involves three distinct steps: \textbf{Prediction}, \textbf{Match}, and \textbf{Update}, each of which is described below.

\begin{enumerate}[leftmargin=*]
    \item \textbf{Prediction:} The RotatedSunFrame framework within $Sunpy$ \citep{sunpy_community2020} is used to describe a rotationally transformed coordinate system within the context of solar coordinates. This tool computes the displacement of each pixel in a solar active region over time, enabling predictions about its future position and configuration. The morphology of active regions is influenced by both local plasma flows and global differential rotation. However, the displacement of the active region caused by local plasma flow is substantially smaller than that caused by differential rotation. Consequently, the effects of differential rotation become more pronounced, thereby enabling the implementation of the predictive method.
    \item \textbf{Match:} The dataset used in our study comprises magnetograms recorded at an interval of $\delta t$, i.e., every 192 minutes or every three hours. After detecting the active regions at time $t$, the time interval allows us to predict their location and morphology of these regions at time $t+\delta t$ using the first step. Simultaneously, we can detect the active regions at time $t+\delta t$ to determine their actual location and morphology, which is referred to as the ground truth. Next, we compute the similarity matrix across all active regions within the \textbf{Prediction} results and the ground truth using Hamming distance \citep{norouzi2012hamming}, a measure of the difference between two equal-length character strings or binary sequences. In this context, after converting each active region and others to binary values of 1 and 0, the Hamming distance is determined by summing the number of points within the region where the resulting value is non-zero after subtracting the \textbf{Prediction} results from the ground truth. Distance is the difference between the \textbf{Prediction} results and the ground truth, therefore, the similarity can be expressed by the following formula:
    \begin{equation}
        Similarity = 1 - \frac{D_{Hamming}}{Size},
        \label{eq: similarity}
    \end{equation}
    where $Size$ is the number of pixels in the active region in the ground truth. 
    
    \item \textbf{Update:} Assuming that $M$ ARs are present at time $t$ and $N$ ARs are present at time $t+\delta t$, an $M \times N$ similarity matrix can be constructed. This matrix helps identify the correspondence between the most similar active regions. Using a predetermined similarity threshold, we can determine whether the ARs from the two different frames are indeed the same. If they are deemed identical, the existing ID number is retained. Otherwise, a new ID is assigned. A memory field is incorporated into the tracking algorithm when specific challenges arise, such as when an object is not detected in one frame but reappears in the next or when the target splits or merges due to the active region evolution or the outcomes of the detection algorithm. This field employs a queue to retain some historical frames, with a default of ten frames. This approach minimizes false and missed tracking over extended durations.
    
\end{enumerate}

\begin{figure}
    \centering
    \includegraphics[width=\linewidth]{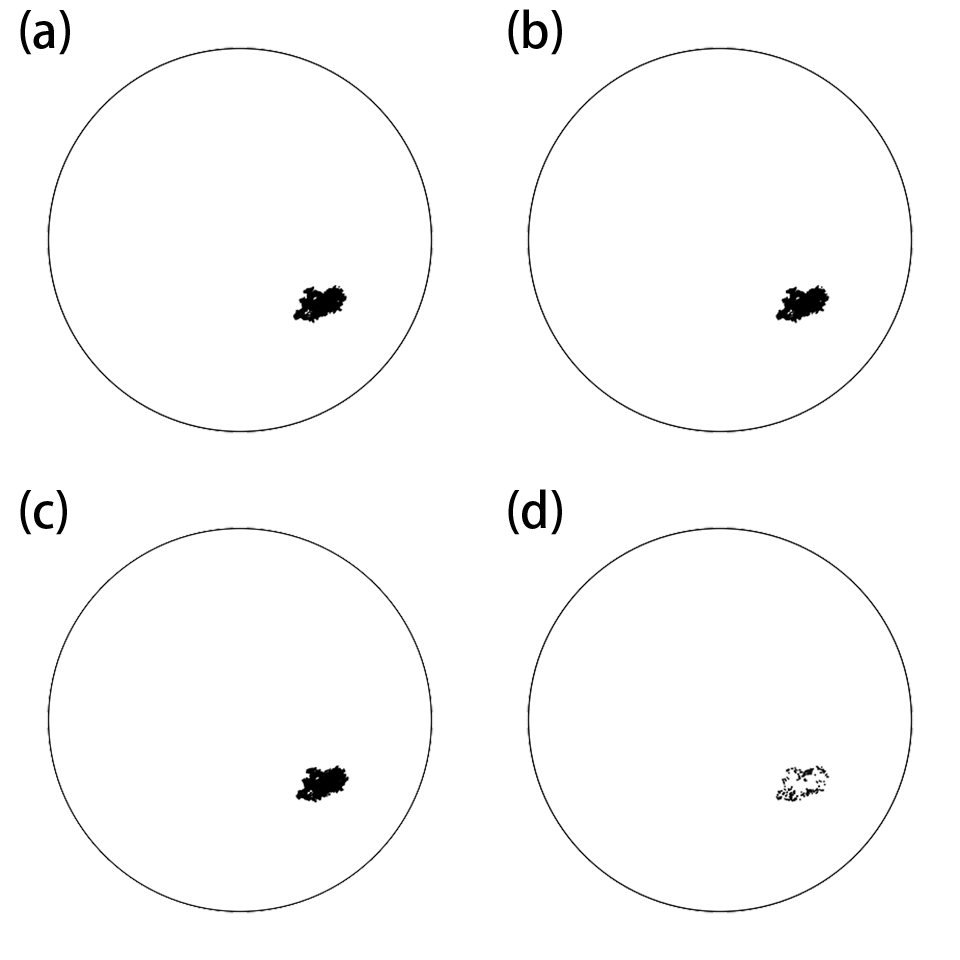}
    \caption{An example shows the entire procedure of \textbf{Prediction} and \textbf{Match} for a specific active region AR2 in Figure~\ref{fig: HDBSCAN}(c), which is shown as the black area in the image. Panel (a) The current position and shape of the AR. Panel (b) The predicted position and shape of the AR at the next moment according to panel (a) and the time interval. Panel (c) The detected position and shape of the AR at the next moment. Panel (d) The difference between panel (b) and panel (c).}
    \label{fig: tracking process}
\end{figure}

\begin{figure*}
    \centering
    \includegraphics[width=\linewidth]{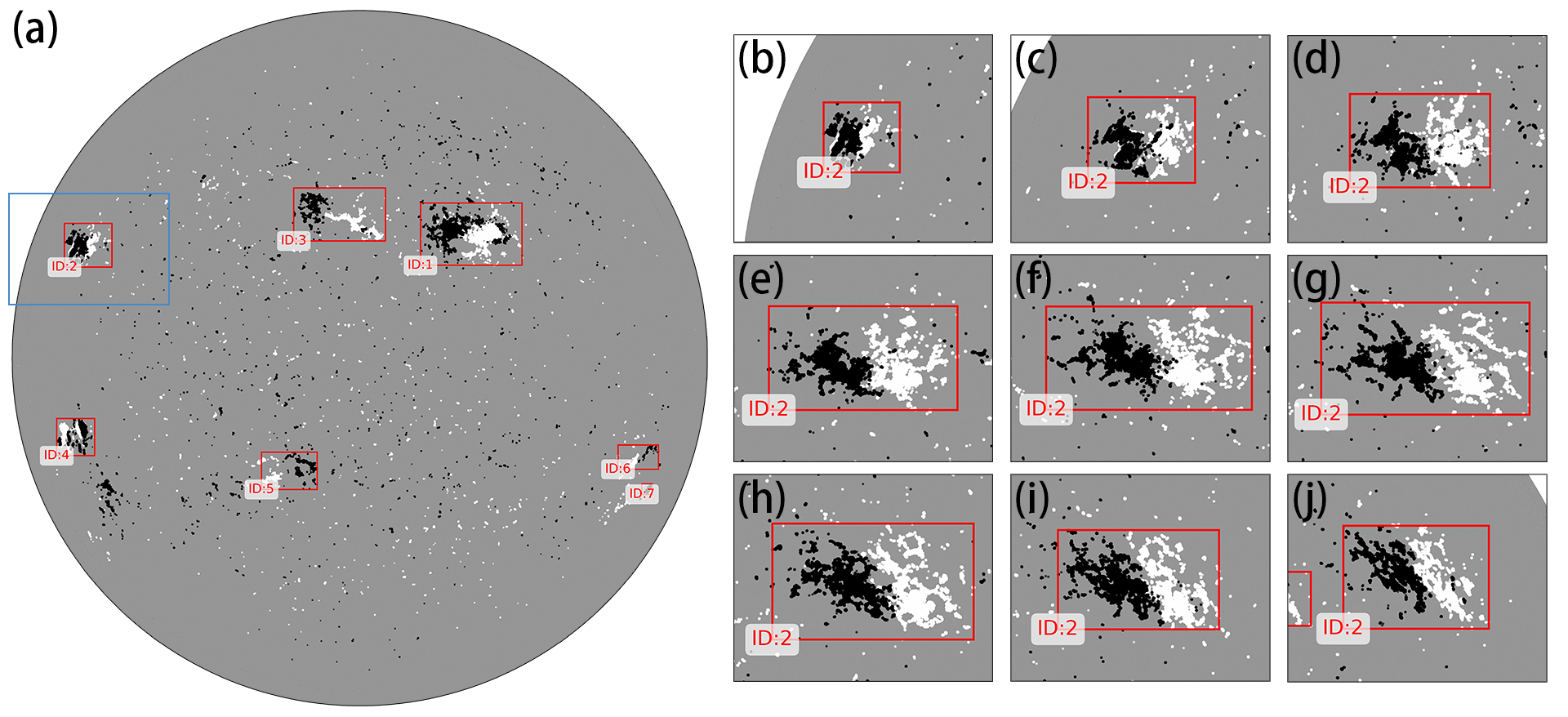}
    \caption{An example of tracking result. Panel (a) The active regions detected by the HARD method on 2022 February 1. Panels (b--j) The trajectory of the active region AR2 from February 1 to February 9, with a time interval of one day.}
    \label{fig: Track}
\end{figure*}

Figure~\ref{fig: tracking process} shows an example to display the entire \textbf{Prediction} and \textbf{Match} process for AR2 in Figure~\ref{fig: HDBSCAN}(c). Figure~\ref{fig: tracking process}(b) is predicted from Figure~\ref{fig: tracking process}(a) via differential rotation over an interval of 192 minutes (since this sample is from the SOHO/MDI magnetograms). Figure~\ref{fig: tracking process}(a) contains $5,227$ pixels. Figure~\ref{fig: tracking process}(c) shows the actual detected AR at $t+\delta t$ (192 minutes), known as the ground truth, and includes $5,154$ pixels. Figure~\ref{fig: tracking process}(d) is derived from the frame difference between Figure~\ref{fig: tracking process}(b) and Figure~\ref{fig: tracking process}(c) and contains $313$ pixels, which also corresponds to the Hamming distance. Finally, the similarity between the AR2 at times $t$ and $t+\delta t$ is calculated to be $0.94$ using Equation~(\ref{eq: similarity}). Figure~\ref{fig: Track} shows the trajectory of AR2 in panel (a) moving from the east to the west. While the morphology of the AR evolves as a result of differential rotation and its inherent dynamics, the tracking method takes advantage of the predictability of differential rotation to ensure that the AR's identifier remains unchanged, as shown in panel (b--j).

\subsection{Performance of the Tracking method} \label{track performance}

The active regions detected by the DSARD and HARD methods in Section~\ref{detection} are tracked using the aforementioned method. The multi-objective tracking evaluation indicators \textit{MOTA}, \textit{IDR} and \textit{R\_MT} \citep{ristani2016performance} are calculated to measure the detection and tracking performance of the model with NOAA labeled as the ground truth (\textit{GT}). The definitions of these indicators are as follows:
\begin{equation}
    MOTA = 1-\frac{FN+ID\_switch}{GT},
    \label{eq: MOTA}
\end{equation}
\begin{equation}
    IDR = \frac{IDTP}{IDTP + IDFN},
    \label{eq: IDR}
\end{equation}
\begin{equation}
    R\_MT = \frac{MT}{GT}.
    \label{eq: R-MT}
\end{equation}

In the multi-object tracking evaluation, \textit{FN} (False Negatives) represents the total count of instances across all frames where a ground truth target is present but is not detected by the method. In contrast, \textit{IDTP} (Identity True Positives) and \textit{IDFN} (Identity False Negatives) are identity-aware metrics. \textit{IDTP} counts the total number of frames across a trajectory where a ground truth identity is correctly and consistently associated with its predicted identity. \textit{IDFN} counts the total number of frames where a ground truth identity is either not detected or is incorrectly associated with a different identity. The term \textit{ID\_switch} indicates that the ID associated with the same NOAA number in the preceding and succeeding frames has changed. \textit{MT} means most tracked target, signifying that no more than two frames within the active region remain untracked. \textit{MOTA} measures the overall accuracy of the tracker by penalizing detection and identity association errors. \textit{IDR} measures the tracker’s ability to correctly recall and maintain target identities. \textit{R\_MT} measures the tracker’s coverage by calculating the proportion of ground truth targets that are successfully tracked for most of their duration. These three metrics are bounded within the interval $[0, 1]$, where larger values denote better tracking performance.

\begin{figure*}
    \centering
    \includegraphics[width=\linewidth]{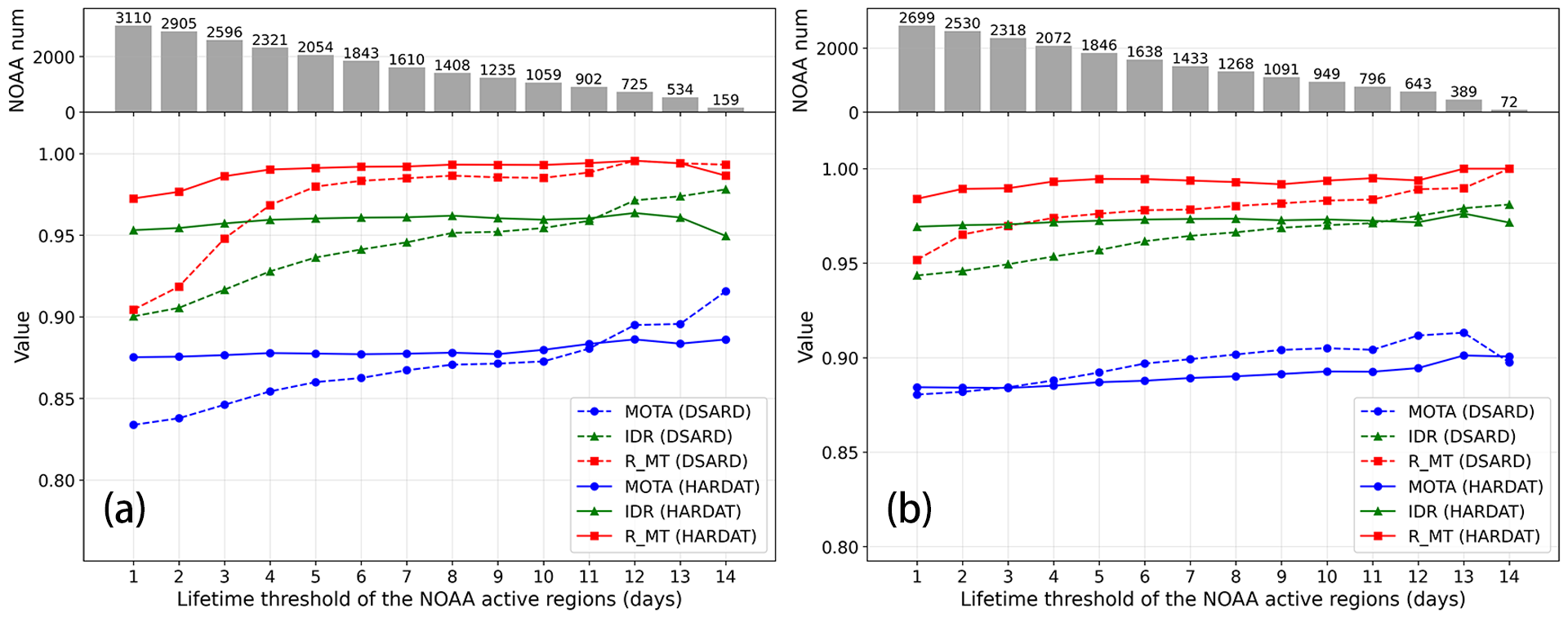}
    \caption{Two line charts of the three tracking evaluation indicators MOTA (blue), IDR (green) and R\_MT (red) as a function of lifetime threshold of the NOAA active regions. The dotted line represents the DSARD method, while the solid line represents the HARD method. The histogram above shows the number of NOAA numbers under the current lifetime threshold. Panel (a) The result of SOHO/MDI magnetograms. Panel (b) The result of SDO/HMI magnetograms.}
    \label{fig: Track_Evaluation}
\end{figure*}

Figure~\ref{fig: Track_Evaluation} illustrates the variation of the three evaluation metrics relative to the lifetime threshold of the NOAA active regions. Due to NOAA's one day temporal resolution, a larger abscissa value indicates a longer duration of visibility of the active region, with a maximum extent of approximately 15 days, corresponding to half of the solar rotation period. The DSARD and HARD methods both consistently demonstrate high evaluation indicator values. The longer the time span, the greater the tracking accuracy. However, when comparing these indicators between the two methods, HARD generally yields superior outcomes in all instances except for the MOTA when examining SDO/HMI magnetograms. Therefore, we can quantitatively infer that HARD is a more effective method than DSARD for detecting solar active regions. We named the method of HDBSCAN based solar active region detection and tracking as HARDAT.

\section{SVC-based Magnetic Polarity Inversion Line Extraction} \label{analysis}

The HARDAT method provides crucial insights into the position, morphology, and magnetic field intensity of an active region. With this information, we can analyze active region magnetograms to identify features such as the magnetic tilt angle and the polarity inversion line (PIL). The PIL is characterized by zero magnetic field strength and separates areas of opposite magnetic polarity within an active region. Since magnetograms contain only positive and negative intensity pixels, extracting the PIL within the active region is equivalent to identifying the boundary between these pixels. We use a support vector classification (SVC;  \citealt{708428}) model to determine a hyperplane that separates positive and negative pole regions within the magnetogram. This approach effectively reformulates the PIL extraction problem as a convex quadratic programming problem.

%\subsection{SVC-based Magnetic Polarity Inversion Line Extraction} \label{SVC-based Magnetic Polarity Inversion Line Extraction}

SVC is a supervised learning algorithm designed for classification. Its objective is to classify new data points into categories by identifying the optimal hyperplane that maximizes the margin between two data classes. Due to the complex and unpredictable morphology and magnetic field distribution in active regions, using kernel techniques is essential to project the data into a high-dimensional feature space and enabling nonlinear classification \citep{shawe2004kernel}. In magnetogram data classification, the data are processed in an up-dimensional manner by using a radial basis function (RBF) kernel, also known as Gaussian kernel. Classification outcomes using the SVC method with an RBF kernel depend on two parameters: the kernel parameter $\gamma$ and the penalty parameter $C$. The penalty parameter $C$ regulates the tolerance of a classifier for misclassification. A higher value of $C$ signifies lower tolerance for inaccuracies, resulting in a narrower soft margin for classification. The parameter $\gamma$ in the RBF kernel function determines the rate at which the kernel function decays. A larger $\gamma$ results in a more rapid decay, which implies only very nearby points are treated as the same class. Based on this concept, selecting a larger $\gamma$ will lead to the classifier producing more intricate and convoluted decision boundaries, while selecting a smaller $\gamma$ will yield a smoother, more linear boundary. After a series of parameter adjustments, $C$ was set to 10 and $\gamma$ to 0.02, considering the PIL extraction results and the generalization ability of the model.  

\begin{figure*}
    \centering
    \includegraphics[width=\linewidth]{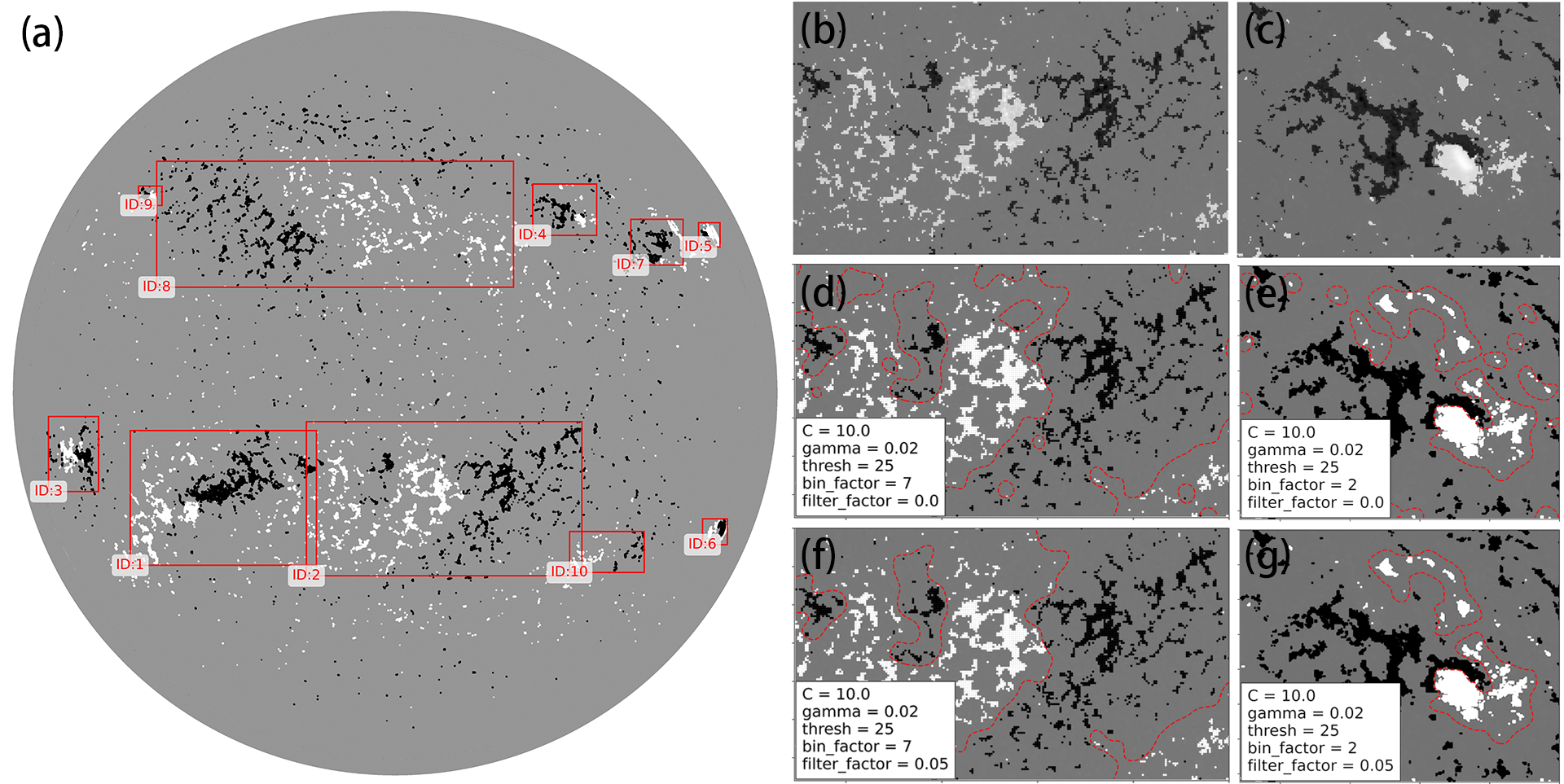}
    \caption{PIL extraction based on the RBF kernel SVC classification. Panel (a) The HARD detection results of the SDO/HMI magnetogram on 2023 February 1. The right two columns represent the PIL extraction steps for the active regions AR2 and AR4, respectively. Panel (b) and panel (c) represent the original image after threshold segmentation. Panel (d) and panel (e) show the extraction result without filtering. Panel (f) and panel (g) show the extraction result with a filter factor of 0.05.}
    \label{fig: svc_PIL_extraction}
\end{figure*}

The process can be divided into four main steps: (1) adaptive downsampling, (2) threshold segmentation, (3) SVC classification, and (4) filtering of small contours. Figure~\ref{fig: svc_PIL_extraction} illustrates the process of extracting two active regions identified by HARD on the same magnetogram. These ARs exhibit distinct differences in size and morphology. Step 1 involves adaptive downsampling, which averages the image based on the size of the active region. This step aims to extract key features and enhance the efficiency of SVC calculations. Step 2 uses threshold segmentation to target lower magnetic field strengths. This ensures that a sufficient number of sample points are available for training, which enhances the accuracy of the results, as shown in Figures~\ref{fig: svc_PIL_extraction}(b) and (c). Step 3 performs SVC classification operations on the images to maintain the decision boundaries, as shown in Figures~\ref{fig: svc_PIL_extraction}(d) and (e). Due to the small threshold established in Step 2, the classification outcomes include an excessive amount of details. This results in the presence of small, independent, closed contours present in the figure. Consequently, Step 4 filters these contours based on the size while preserving the primary decision boundary as the PIL extraction outcome.

\section{Discussion and Conclusion} \label{discussion and conclusion}

We presented an improved automated technique based on the DSARD for detecting and tracking solar active regions using the LOS full-disk magnetograms from both SOHO/MDI and SDO/HMI datasets. Unlike the DSARD method, the proposed technique incorporates two additional parameters, $\beta$ and $\gamma$, which are determined by the cluster size. These parameters improve the cluster integration and subsequently provide a clearer AR structure. Due to the fixed density threshold characteristic of DBSCAN, we used an advanced clustering algorithm called HDBSCAN. HDBSCAN can group data points based on cluster stability to achieve dynamic density clustering, ultimately improving the generalization capability of the model.

We sampled the LOS full-disk magnetograms of SOHO/MDI from 1996 to 2011 and SDO/HMI from 2010 to 2024 at time intervals of 192 minutes and three hours, respectively. Then, we implemented the active region detection using two methods: DSARD and HARDAT. Since the DSARD method does not incorporate a temporal tracking algorithm for active regions, we have independently tracked the two aforementioned detection outcomes. This tracking approach employs solar differential rotation to predict the position and shape of active regions in the next time step. Next, we computed the Hamming distance between the predictions and the actual active regions to derive a similarity matrix to update the active region identifiers for tracking purposes. We used the NOAA catalog as the ground truth to evaluate the DSARD and HARDAT test and tracking outcomes, respectively. This evaluation was quantified using three parameters: \textit{R\_MT}, \textit{IDR}, and \textit{MOTA}. We found that both DSARD and HARDAT performed well in all three metrics, though HARDAT performed better overall.

\begin{figure}
    \centering
    \includegraphics[width=\linewidth]{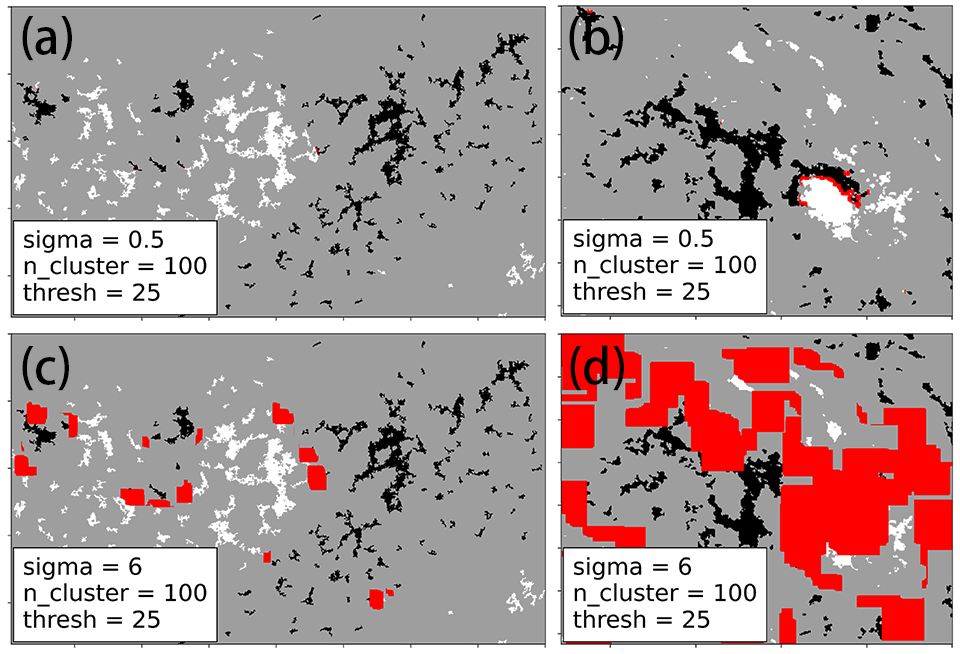}
    \caption{Results of PIL-mask extraction using different sigma and n\_cluster parameters are presented, with the red areas denoting the extracted PIL regions. The active region selected is the same as that shown in Figure~\ref{fig: svc_PIL_extraction}.}
    \label{fig: mask_PIL_extraction}
\end{figure}

\cite{ran2022relationship} extracted the active region PILs using the mask method, referred to hereafter as PIL-mask, as shown in Figure~\ref{fig: mask_PIL_extraction}. The PIL-mask method primarily involves applying DBSCAN clustering to the target active region again, identifying the larger clusters as the main components, and then applying Gaussian smoothing to the area to expand the cluster boundaries. This process ensures that the positive and negative regions overlap to some degree, with the overlapping area ultimately designated as the PIL region. However, as Figure~\ref{fig: mask_PIL_extraction}(a--d) shows, the shape of the active region is often intricate, with variable distances between these clusters. Using a larger sigma— which determines the smoothing radius in Gaussian filtering — causes the mask to indiscriminately encompass the entire local area when the positive and negative regions are close together. Thus, the PIL-mask approach has inherent limitations when it comes to extract the PIL in active regions. 

\begin{figure*}
    \centering
    \includegraphics[width=\linewidth]{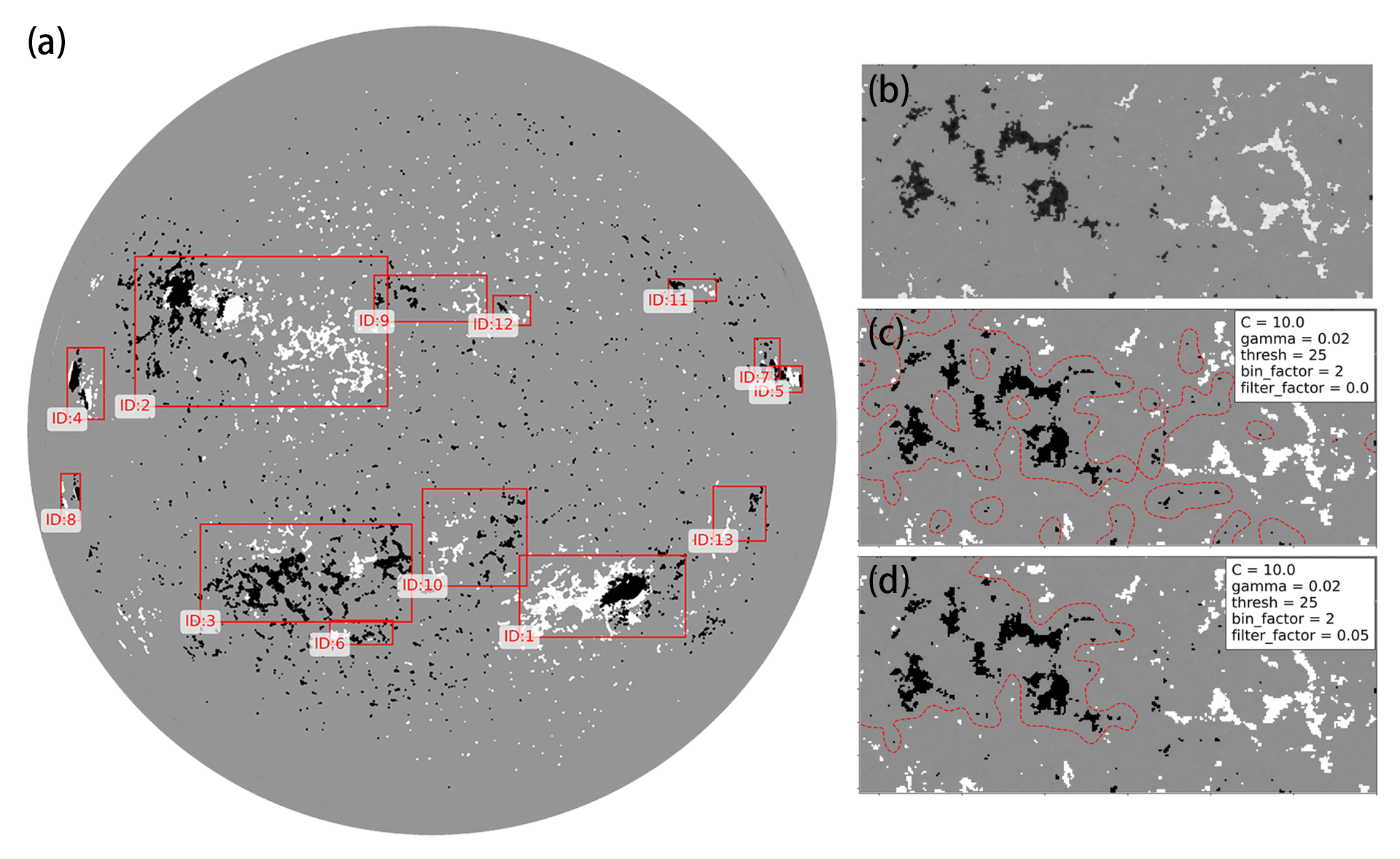}
    \caption{An example of PIL-svc extraction on July 15, 2023. Panel (a) The HARD detection results. Panel (b--d) are similar to Figure~\ref{fig: svc_PIL_extraction} for the PIL extraction process, but for a decayed active region AR9.}
    \label{fig: Fig_10}
\end{figure*}

Therefore, we proposed a novel method for PIL extraction that relies on SVC, termed PIL-svc. This approach primarily leverages the classification principles of SVC by employing the decision boundary to separate the positive and negative poles of the active region. Due to the complex structure of active regions, we employ an RBF kernel technique to increase the dimension of the data and help identify nonlinear decision boundaries. Although PIL-svc may not perform as well as PIL-mask under optimal conditions, it exhibits superior generalization capabilities. This strengthens its ability to efficiently handle the batch processing of neutral line extraction tasks within active regions. Consequently, PIL-svc is more effective than PIL-mask in long-term statistical analysis. Additionally, since the SVC classifier can produce the decision boundaries based on discrete distributions, PIL-svc achieves good extraction results in decayed or diffuse active regions with a PIL structure, such as AR9 shown in Figure~\ref{fig: Fig_10}. Each panel in Figure~\ref{fig: Fig_10} is similar to Figure~\ref{fig: svc_PIL_extraction} for the PIL extraction process.

In summary, this study developed and applied improved automated techniques to detect, track, and analyze solar active regions using LOS full-disk magnetograms from SOHO/MDI and SDO/HMI. The validity of these methods was evaluated by comparing them with the NOAA catalogs. The key methodological advancements and their resulting outcomes are summarized below:

\begin{enumerate}
    \item The HDBSCAN-based Solar Active Region Detection and Tracking (HARDAT) method utilizes dynamic density clustering based on cluster stability to demonstrate superior capability in identifying smaller and more diffuse magnetic structures while maintaining the morphological integrity of large, complex ARs. The HARDAT method provides more generalized detection performance across diverse solar magnetograms with varying spatial resolutions.

    \item The solar differential rotation based tracking algorithm significantly reduces identity switches and tracking fragmentation by integrating physical models of solar surface motions for prediction and using Hamming distance-based similarity metrics for data association. This method can reliably handle AR evolution events, such as splitting, merging, and migration, across extended temporal sequences.

    \item The novel PIL extraction method redefines the identification of polarity inversion lines as a classification task. It accomplishes this using support vector classification with a radial basis function kernel. This method can generate precise and adaptable decision boundaries that effectively distinguish opposite magnetic polarities within complex structures of active regions.
\end{enumerate}

\begin{acknowledgements}
We are grateful to the SOHO and SDO teams for providing the observational data. The present work was supported by National Natural Science Foundation of China (NSFC) under grants 12173019, 12333009, and 12127901, National Key Research and Development Program of China (2020YFC2201200), CNSA project D050101, the Fundamental Research Funds for the Central Universities 14380065, KG202506, and the Young Data Scientist Program of the China National Astronomical Data Center, as well as the AI \& AI for Science Project of Nanjing University. 

\end{acknowledgements}

\bibliography{reference}{}
\bibliographystyle{aasjournal}

\appendix
\section{HDBSCAN Processing Logic for Solar Magnetograms} \label{HDBSCAN flowchart}
\begin{figure*}
    \centering
    \includegraphics[width=.9\linewidth]{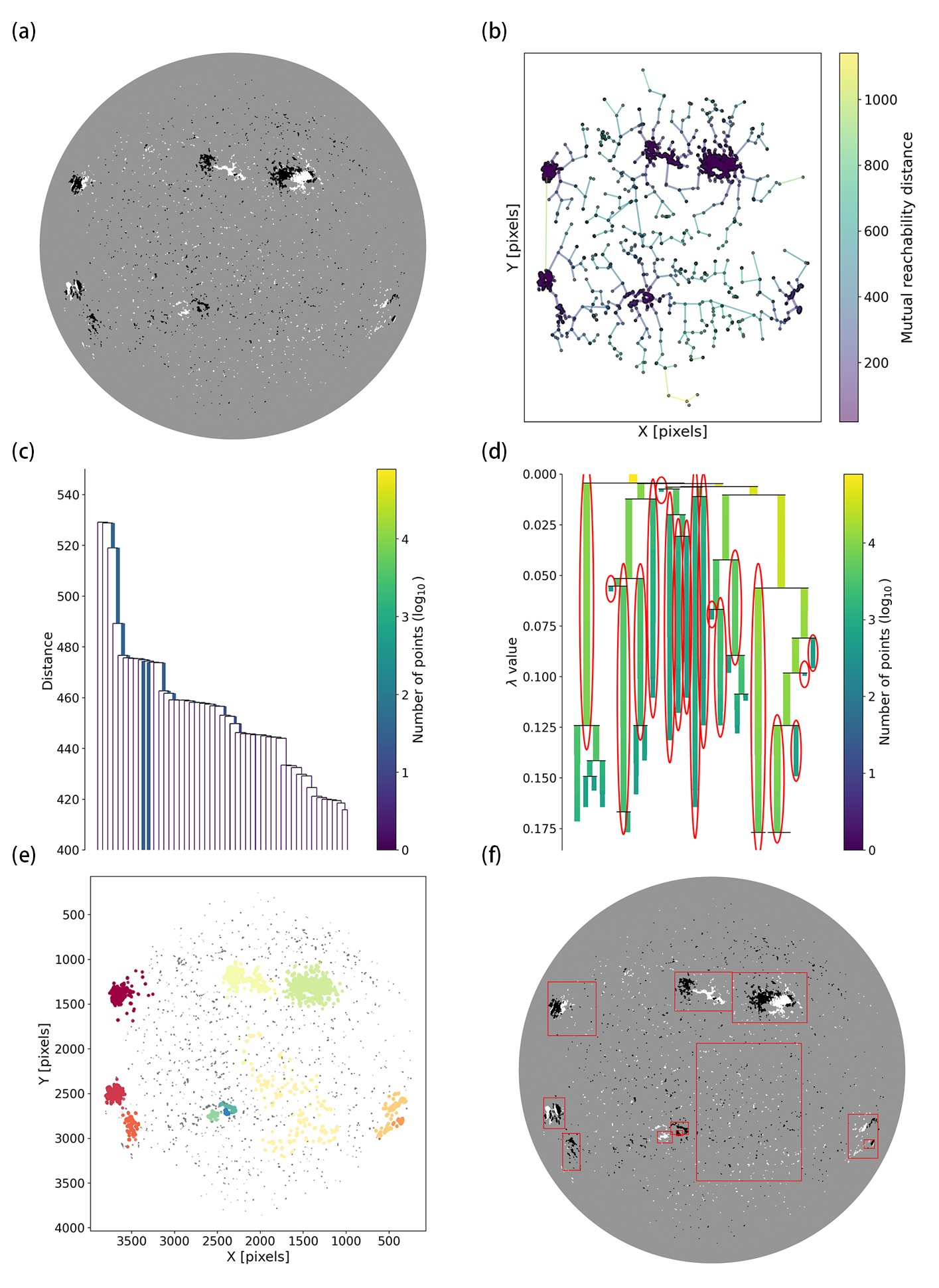}
    \caption{The flowchart of the HDBSCAN clustering applied to solar magnetograms on 2023 February 1. Panel (a) The result after threshold segmentation. Panel (b) Minimum spanning tree construction with random 5000 pixels from (a). Panel (c) Single linkage hierarchy. Panel (d) Condensed hierarchy tree. Panel (e) The HDBSCAN clustering result. Panel (f) The clusters with bounding box and the background.}
    \label{fig: HDBSCAN clustering flowchart}
\end{figure*}

In Section~\ref{HARD}, we used HDBSCAN to achieve dynamic density clustering.  Figure~\ref{fig: HDBSCAN clustering flowchart} shows the HDBSCAN clustering flowchart on solar magnetograms. Figure~\ref{fig: HDBSCAN clustering flowchart}(a) shows the seed pixels after the magnetic field strength threshold segmentation. Subsequently, the mutual reachability distance is computed for all point pairs using the core distance defined by $minSamples\_H$, and a minimum spanning tree (MST) is constructed based on these distances, as shown in Figure~\ref{fig: HDBSCAN clustering flowchart}(b). This MST is then converted into a single linkage tree to represent the complete hierarchy of connected components, as shown in Figure~\ref{fig: HDBSCAN clustering flowchart}(c). Given the large number of points, the plot parameter $truncate\_mode$ is set to ``lastp'' with $p=50$ to simplify the tree for improved visual clarity. 

To further simplify this complex structure, we transform the single linkage tree into a condensed tree by filtering out clusters smaller than the $minClusterSize$ parameter, as shown in Figure~\ref{fig: HDBSCAN clustering flowchart}(d). To find the most stable clusters, the algorithm calculates the stability of each branch by summing the persistence of its points (defined by the inverse distance $\lambda$) and selects the most persistent branches based on these calculated stabilities. In Figure~\ref{fig: HDBSCAN clustering flowchart}(d), the red ellipses indicate the most stable clusters selected by the algorithm. The $\epsilon\_ClusterSelection$ then acts as a final decision rule to determine whether to merge these clusters.
 
 The extracted clusters are visualized in Figure~\ref{fig: HDBSCAN clustering flowchart}(e), where different colors represent distinct identified clusters, while gray points represent background noise. Finally, the results are mapped back onto the original solar magnetograms with bounding boxes, as shown in Figure~\ref{fig: HDBSCAN clustering flowchart}(f).

\section{Parameter Beta and Gamma Analysis of HARDAT} \label{param_analysis}
\begin{figure*}
    \centering
    \includegraphics[width=0.8\linewidth]{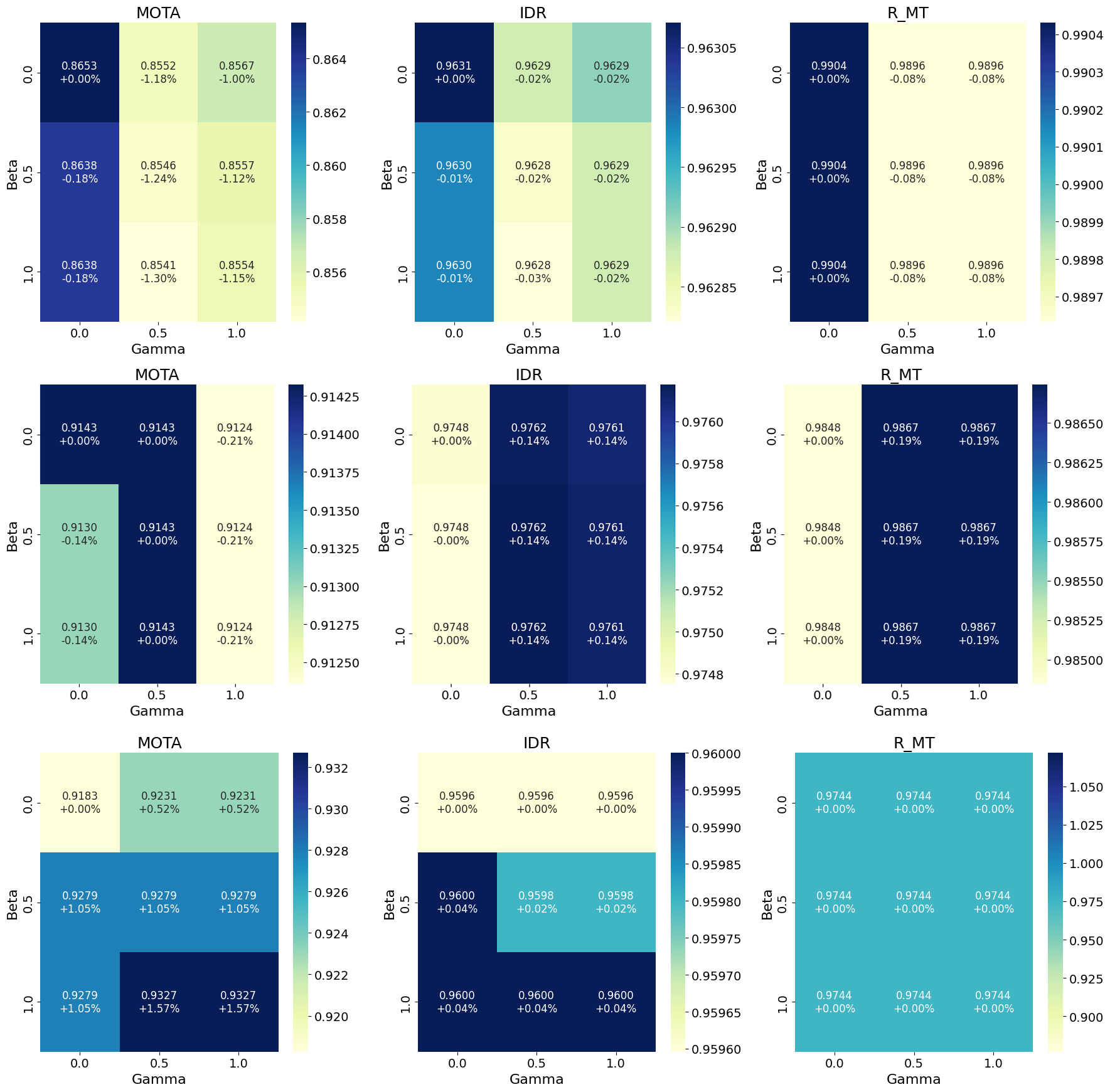}
    \caption{Thermogram of the different evaluation indicators changes depending on the parameters $\beta$ and $\gamma$. Take the lifetime threshold of the NOAA active regions to be 2. The first row corresponds to 2000--2002 (the maximum of solar cycle 23), the second row corresponds to 2003--2005 (the decay phase of solar cycle 23), and the third row corresponds to 2007--2009 (the minimum of solar cycle 23).}
    \label{fig: Beta_Gamma}
\end{figure*}

In Section~\ref{detection}, we introduced the $\beta$ and $\gamma$ parameters to improve the completeness of the active region morphology in the detection process. Here, we analyze how the selection of these two parameters influences the detection results. The evaluation criteria are determined by the three evaluation parameters in Section~\ref{track}. Three time periods were selected that corresponding to the different stages of a solar cycle: 2000--2002 (the maximum of solar cycle 23), 2003--2005 (the decay phase of solar cycle 23) and 2007--2009 (the minimum of solar cycle 23), respectively. We selected nine cases of the values 0, 0.5, and 1 for $\beta$ and $\gamma$, respectively, to calculate the values of \textit{MOTA}, \textit{IDR} and \textit{R\_MT} in different parameters and different periods, as shown in Figure~\ref{fig: Beta_Gamma}. It is evident that the parameters $\beta$ and $\gamma$ exert minimal influence on the detection and tracking outcomes, primarily affecting the ultimate morphology of the active region. Therefore, for the purpose of calculating efficiency in statistical analysis, we assign both $\beta$ and $\gamma$ a value of 0.

\end{document}